# Modeling skier behavior for planning and management. Dynaski, an agent-based in congested ski-areas

Alexis Poulhès (corresponding author), Ecole des Ponts, Laboratoire Ville Mobilité Transport (UMR Ecole des Ponts ParisTech, Université Gustave Eiffel), 14-20 boulevard Newton - Cité Descartes - Champs-sur-Marne, 77447 Marne-la-Vallée Cedex 2, France. alexis.poulhes@enpc.fr

Paul Mirial, Ecole des Ponts, Laboratoire Ville Mobilité Transport (UMR Ecole des Ponts ParisTech, Université Gustave Eiffel), 14-20 boulevard Newton - Cité Descartes - Champs-sur-Marne, 77447 Marne-la-Vallée Cedex 2, France

## Abstract
In leisure spaces, particularly theme parks and museums, researchers and managers have long been using simulation tools to tackle the big issue associated with attractiveness – flow management. In this research, we present the management and planning perspective of a multi-agent simulation tool which models the behavior of skiers in a ski-area. This is the first tool able to simulate and compare management and planning scenarios as well as their impacts on the comfort of skiers, in particular ski-area waiting times. This paper aims to integrate multiple data source to calibrate the simulation on a real case study. An original field survey of users during a week details the skier population. The first average skier speeds are calculated from GPS data on one ordinary day. The validation data are used to calibrate the parameters of the behavioral model. A demonstration of the simulation tool is conducted on the La Plagne ski-area, one of the largest in France. A test case, the construction of new housing in a station near the ski-area, is conducted. An addition of 1620 new skiers delays the skier average waiting time by 12%.

## Introduction

According to the world's leading expert on ski-areas, Vanat (2019), there are 2,113 ski resorts worldwide with 26,334 ski-lifts. While in some developing countries, such as China, ski-areas are being created, the opposite phenomenon is taking place in traditional skiing countries, where some small low-lying stations are closing as a result of decreasing snow cover and the excessive cost of installing snow cannons (Bachimon 2019, Damm et al. 2014). In France or in the United States, however, these closures do not benefit the large, high-altitude stations, where the number of annual skier days has stopped rising (Domaine Skiable de France, 2014, National Ski-areas Association 2019). Holidaymakers are changing their behavior patterns, looking for a variety of activities. At the same time, to compensate for the lack of snow, resorts have diversified the range of activities they offer, moving away from an exclusive focus on skiing (Hewer and Gough 2018, Perrin-Malterre 2015, Bourdeau 2009). However, unlike other destinations, France relied solely on skiing forgetting other activities (Odit France 2005). The decline in attractiveness linked with changes in the image of winter sports resorts that have been built in preservation areas forces ski-areas to limit their renewal efforts to the spaces assigned to them (Needham et Little, 2013). The need for significant investment and the need for staff oblige ski-areas to charge high prices, which also limits their attractiveness (Le Monde 2018, Skiinfo 2015) and their productivity (Goncalves 2013). At the same time, the digital revolution has been rapid and successful in this sector (Gérard 2016, Humair 2017). For customers, digital technology has entered the world of winter sports through the digitisation of ski passes. In some ski-areas, there are now mobile applications that can give users real-time information about waiting conditions, and provide itinerary tools or even snow sports statistics (example of the YUGE application in the Paradiski-area). In terms of ski-area management, snow grooming and snowmaking have entered the era of real-time data. However, there are not yet any forward simulation tools for ski lift management and long-term ski-area planning, as exist for more traditional passenger transportation (PTV VISUM, Citilabs CUBE).

There are very few studies that link simulation approaches to mobility in a ski-area. A theoretical study (which technically could be operational in practice) carried out at the Serre-Chevalier station tried to clarify the evolution of ski-area infrastructure through a fractal approach (Brissaud, 2007). According to the authors, whatever the size of the ski area, the distribution of ski lifts and slopes would follow a fractal geometry. A Swiss project, Juste-Neige, aimed to predict snow thickness using an agent-based approach (Revilloud et al. 2011). The model is based mainly on turnstile counts, with skiers being generated according to changes in ski-lift counts. The agents choose slopes that match the numbers recorded by each count. Tino et al. 2014 propose a queue model to calculate ski lift waiting time, without taking account of any vacant seats caused by the composition of the groups. An internship report (Tharnish et Recla) presents an infrastructure optimization model, with a methodology based on the maximization of the theoretical capacity of the ski-area and the speed of the lifts. Unfortunately, the model is very theoretical and the time complexity is exponential. In another piece of research, (Liu et al., 2010) are building a mechanical simulation model of a group of skiers on an off-piste slope to better determine the probabilities of being present during an avalanche and thereby to assist rescue workers. A leader chooses the direction and each skier independently builds his or her own track in interaction with the leader's track.

In an urban transit network where supply is characterized by a flow over an arc as in ski-areas, multi-agent models are successfully used to define activity days for a synthetic population. However, the aim of the demand assigned to the transit network is only to minimize a cost, in which time is the main component. MATSIM (Horni et al. 2016) is the most successful and widely used model today.

The increasing number of studies that focus on theme parks are closest to the simulation of skiers in an area. The problem of waiting time is the same and customer travel is not governed by the minimization of a cost or time, but rather by the optimization of satisfaction or utility. Different types of mobility models are used. For a general state of the art, the reader is referred to several literature reviews (Camp et al., 2002; Härri et al., 2009). In addition to basic random models (random way point, random walk, or Markovian mobility models), which are unrealistic for simulating human behavior (Rhee et al. 2011, Solmaz et al. 2012), other more sophisticated models are used. Some use GPS tracks (Cheng et al. 2013) to estimate the probability of movement from one area of the park to another area following a decision tree. Others incorporate physical effects, synthetic behavioral models. Vukadinovic et al. (2011) from the Walt Disney Company developed a microsimulation based model to manage flows in theme park environments. Destination choices are calibrated with GPS data but no specific choice behavior is modelling to simulate new activity or long-term planning. The focus of the model developed by Chu et al (2014), in particular, is waiting time. Each attraction has a popularity, a capacity, and therefore a waiting time that depends on the total number of visitors. Solmaz et al. (2015) simulate the evacuation of all customers from a park in the event of a disaster. The behavioral objective is to reach the nearest exit, the physical interactions between agents are taken into account. The choice of destination is also central to the models. In addition to the random way point, Kataoka et al. (2005) represent the theme park in the form of a graph, the probability of visiting each attraction is set arbitrarily. Solmaz et al. (2012) propose a mobility model for a theme park where the choice of destination depends on the waiting time for the attraction obtained from GPS data. The longer the wait, the more customers are attracted. This type of model cannot be used in the case of scenarios involving changes in the supply. Hsu et al. (2005) simulate the probability of moving from one location to another on a campus and the time spent on each location from a field survey. Another study also uses an ordered logit model to predict the waiting time at each attraction in a theme park from a survey (Kemperman et al. 2003). GPS data have become an important source for understand behaviors to highlight sequences as Kemperman et al. (2004). In particular, they distinguish the activity pattern of the first-time visitors and the repeaters. Drawing on the theory of time geography, Biremboim et al. (2013) analyze people's behavior in a theme park, Portaventura in Spain. They determine stable pattern of visitors regardless the whether the season congestion or opening hours.

In this article, we detail a simulation tool which responds to a large number of management and planning situations for the ski-area operators. The user will be able to compare different infrastructure proposals and their impacts on flows. The objective is thus to provide a model that can be used for predictive purposes, which takes into account the characteristics of the ski area and the behavior of skiers so that different solutions for change in the ski area can be evaluated. A precedent research exposed the multi-agent model as a theoretical way without a real background. The produced results are then based on several original data essential for the model. This diversified sources of data improve the understanding of the skier behavior in a ski-area. As an intermediate objective, the skier speed calculation from GPS data provide travel time on the slopes for the model.

A ski area is a dedicated space where skiers can delight the experience of descending snow slopes. To regain altitude before each ski run, they use ski lifts. The slopes and ski lifts form a network of possible routes. Lines of skiers may appear to take certain lifts. Skiers either stay in resorts directly in the ski-area or access the area by bus or car from accommodation further away.

A behavioral model simulate the changing flows of skiers in the ski area in case of management or planning scenario. Skiers change their behavioral patterns in the course of a ski day, which requires each of these "phases" to have its own behavioral model (ski periods, lunch break, return to the resort). Management difficulties for the ski-area operator are maximal with high congestion and planning decision aims to limit this congestion. Hence, the simulation period is a day in the winter holiday when the number of skiers is maximal. In a consistent manner, the used data correspond to the same period. A skier survey is conducted in a specific ski-area, La Plagne, to confirm the behavioral model choices and to build the population structure. The size of groups and the level of ski are determinant in the model. The skier motion in the ski-area is often constraint by the lift queues or by the lifts. The other time on the slope depend on the speed of each skier. To the best of our knowledge, neither study calculate an average speed on the slope. In a consistent way, the GPS data of the same ski-area, La Plagne, are used to perform speed depending on the level of the skiers and the slope difficulty. Finally, behavioral parameters are calibrated with the validation data of La Plagne.

The main contributions of this article are as follows:

(i) Detail of a list of possible operational uses of this type of modelling tool

(ii) Original survey data produced in the course of this research to calibrate skier population

(iii) Using GPS data provided by the ski-area which saves the tracks of skiers who use the ski-area's smartphone app to evaluate skier speeds

(iv) Calibration of the model with validation data from all the ski lifts in the ski-area over one month during French school holidays (the busiest period).

The first part of the article describes the different operational uses of this type of tool. The second part presents the multi-agent model for skiers moving in the ski-area. The different calibration steps — estimation of skier speeds, composition of the skier population, and optimization of the model parameters on the validation data — are presented in part 3. The calibration results presented all relate to the La Plagne ski-area. The last part of the paper illustrates an operational use of the model on the La Plagne ski-area, with the construction of new housing in one of the resorts attached to the area and its impact on all skiers.

# 1 Scope of application

The aim of this first part is to introduce this new field of research, the modelling of skiers in a ski-area in an operational context. It describes a number of applications for both planning and management, and introduces the theoretical model and the methodological choices. It therefore explains the operational need for this type of modelling.

## 1.1 The ski-area

A ski-area is an open space in mountains that is mainly active in winter. It contains designated snow slopes for skiers and lifts to carry skiers to higher altitudes. Ski-area companies or operators manage the area and the infrastructure under public-service contracts. They do not own the land but are responsible for planning and management.

Innovative data collection solutions are widespread in this field of activity. In the last 20 years, it has become possible to access ski lifts using a single digital pass in many ski-area. Data is collected for each ski lift. Major ski stations offer skiers smartphone applications with multiple tools to improve their ski experience. Many ski-areas use these data to identify the sociological profiles of users and to track GPS itineraries. Startups continue to offer a new data collection solutions, many of which can generate graphical control panels.

A number of books and articles have been written on what makes a winter sports resort attractive (Konu et al., 2011). Few, however, have focused on the ski-area itself. Nevertheless, the factors that determine the attractiveness of a ski-area are specific: snow cover, slope exposure, weather, but also congestion. While snow quality and depth are major issues for ski-areas (Spandre et al. 2016) and are a major target of investment, excessive crowding during certain periods of the skiing season seems to be a factor that Ski lift companies cannot control. However, optimizing waiting time is a key priority for operators. In satisfaction surveys, waiting time at ski lifts is a dominant preoccupation (Hummel, 2019).

Several studies have looked into strategies that are effective in attracting skiers. The size of the ski-area and the lack of on-ski connections between ski resorts are key reasons for the failure of ski-lift companies (Falk 2013, Needham & Little 2013). Long-term projections predict a reduction in skiable areas and opening days, despite the increasing number of snow-making installations (Damm 2014).

Ski-area operators are therefore now trying to improve the quality of skiers' experience by means of field tools. A number of experiments and tests have been conducted to limit waiting time. In Switzerland, skiers can pay an extra charge to jump in line. In some ski stations, individual skiers can wait in a separate line to fill a space in a chair.

On the slopes, high skier density can force skiers to slow down and then feel unsafe, reducing the attractiveness of the experience. Better flow distribution could reduce traffic peaks on certain slopes. In large ski-areas, skiers can be kept informed about parts of the ski-area that are less congested by means of display panels. The most effective measure is to replace an old ski lift with one that has greater capacity. However, a ski-area is complex and if all ski lifts are optimized, skiers will spend less time actually on the lifts and more time on the slopes and in the lift lines.

By modelling the behavior of skiers as they move around, it is possible to test any modification — even major modifications — in the ski-area. Such a modification might be temporary, such as a lift closure or failure, but might also be more structural, and therefore require modelling of the potential long-term effects.

## 1.2 Modelling for planning

*Upgrades*

At present, ski lift upgrade planning is simply based on expert advice and the diagnosis of attendance from historical data and experience. However, a lift is a big investment for a ski-area, one that will be in place for more than 30 years. A simulation tool is a way to compare upgrade priorities, priorities for routes or lift types, defined by their speed, capacity, or ease of use. The choice of new slopes or changes in the layout of a slope, though more restricted by topology, can also be assessed through simulations.

The effect on skier numbers of joining 2 ski-areas together can also be simulated and anticipated by means of such a tool.

*Ski-area creation or extension*

In mountainous and snowy countries that still have very few ski-areas, such as China, a simulation tool can help in planning the size of the area and the infrastructures according to anticipated use and local constraints. Extensions to a ski-area can also be guided by simulation. Operators can thus test the impact of a range of choices on skier practices and congestion levels. Slope layout, ski lifts, and ski lift connections and characteristics, are the primary instruments of action.

*Ski resort urbanization*

Although ski-areas in Europe are mature and currently developing very little, there is still extensive housing construction in the ski resorts that depend on them. Ski resorts like Val-Thorens adapt their total lift capacity to local accommodation levels: "the hourly ski-lift capacity should be twice the accommodation capacity." However, planners do not have the fine-grained spatial detail and precise indicators needed to measure the link between the ski-area's capacity in skiers and the capacity of the resorts in number of beds. The knowledge that a built-up area will correspond to a given number of lodgings and therefore to a given number of skiers per day makes it possible to simulate the impact of a new building or residence on the performance of the ski-area and in particular on congestion. With this kind of simulation tool, therefore, new construction can be targeted on one resort rather than another in large ski-areas where several resorts are linked. But such a tool can also be used to limit the number of new buildings or to promote parallel development of resorts and ski-areas.

*Anticipate climate change or a decline of the ski tourism*

There are many uncertainties about the future of the ski resorts. The decrease in the number of days with sufficient snow depths to ski, the limitation of the size of the open area due to the lack of snow or even changes in the practice of tourists who ski are real risks that the ski areas must anticipate. This tool can then be used to simulate a drop in demand and the adaptation that can be made in the domain. Conversely, if a large part of the domain is closed due to lack of snow, how to prepare the influx of skiers to the rest of the domain.

## 1.3 Modelling for management

While this kind of simulation tool is more suitable for planning with evaluations of typical days and maximum congestion conditions, it can also simulate some typical management cases. Here, the aim will not be to compare scenarios for changes in ski-area infrastructure, but to analyze a scenario of service deterioration. On the basis of the simulation results, managers will thus have to define acceptability thresholds for skier waiting times and decide on operational solutions accordingly. In concrete terms, all that needs to be done in the simulation is to deactivate slopes and lifts and define the length of closure for the day simulated. By proactively testing such simulations, managers can anticipate a set of critical situations.

*Closing a slope or a lift*

When a lift breaks down for part of the day or when a slope is closed due to lack of snow, the impact on the rest of the ski-area can be anticipated by the simulation and even used to inform skiers of potential difficulties. For example, a decision could be made to increase the number of lift staff in certain areas.

*Closing an area*

By extension, an entire sector may be closed if it is served by a single lift which fails, or if there is a shortage of snow in low-altitude or sunlit sectors.

*Testing the closure of lifts or sectors*

In order to limit staff numbers, ski-areas frequently close lifts or parts of their ski-area. The impact and scale of these one-off or longer closures can be evaluated with a simulation tool. The impact on skiers and the increase in waiting time are quantifiable.

*Rationalization of provision in the event of a shortage of skiers*

A fall in the number of skiers on certain days or at certain times of the year may prompt the station manager to close certain lifts to avoid paying too many staff. These closures could be evaluated and optimized to limit the impact on skiers.

*Changes in skier behavior or structure*

For the purposes of long-term management, the risks of change need to be anticipated. In particular, there could be radical changes in skier behavior or in the structuring of skiers into level groups. So simulating different populations of skiers with a much larger proportion of expert or solo skiers would make it possible to see whether the operation of the ski-area is robust to such changes in demand. Similarly, much faster skiers could have a negative impact on average congestion across the ski-area.

## 2 Modelling framework

### 2.1 Representation of the network

A ski-area is a closed space made up of slopes and ski lifts. Any given point is accessible from another point. This means that the ski-area can be perceived as a connected and directional graph. We represent the ski-area as follows: $D = (P \cup R, N)$ is a set of arcs $P \cup R$, in which $P$ is a set of slopes and $R$ a set of ski lifts and a set of nodes $N$. The principle of skiing may be defined as two elements: riding the slopes and regaining altitude on the ski lift. The slopes and ski lifts move in only one direction. The nodes are either points of direction or entry points. Entry points are flow generators such as stations or parking lots beside some of the ski lifts. We will see below that choice points can also be restaurants or destination points.

#### 2.1.1 The slopes

In all ski-areas, slopes are divided into four categories, depending on the skiers' level and abilities, from green (the easiest) to black (the most difficult). Sometimes, slopes cross other slopes. In order to simplify modeling, we have divided all the slopes at each intersection point. Thus, in the graph, the slope representation is a linear arc with a specific length and area, as well as specific features such as snow cover, sun exposure, for example. We assume, in our model, that all skiers use the slopes. So, we have not taken off-piste skiing into account.

### 2.1.2 The ski lifts

Skiers use ski lifts in order to get back to the original altitude they were at before going down the slope. Several types of ski lift meet the various requirements of the ski-area. All ski lifts work in the same way: a steel cable pulls cabins, seats or other items with constant headway along a linear itinerary through the mountain. Speed, capacity and arrival altitude are key factors in the choice of ski lift for the skiers. The most common ones are the chairlift, the cableway and the surface lift. The French website http://remontees-mecaniques.net/ offers a precise explanation of the differences between the types of ski-lifts and their respective functions. In our model, we only focus on the capacity, frequency and speed[1] of the ski lifts. A difficulty parameter makes it possible to introduce differences between technical solutions.

## 2.2 The skier population

Large ski-areas are linked to several ski resorts. Skiers come directly from these ski resorts or from the valley below, by car, and park in a parking lot near a ski lift. These are the skier generators for the ski-area, represented as a node. During the simulation, the departure points are saved for each skier. The model associates the skiers to their origin point – the point where they return at the end of the day. Virtual arcs are created to connect the network and the demand introduced in the skier generators.

We can assess the number of skiers from the number of beds, and the size of the apartments in the ski resorts from the number of parking spaces in the lots. The population structure could be known with surveys, which are currently missing. As for the first validation turnstile data, it can be used as an intermediate demand data. Several types of ski passes are on sale in the large ski-areas. They allow access to different area sizes within the ski-area. Skiers are constrained by their ski passes to stay within a confined zone. In short, we assume skiers can access the whole ski-area.

### 2.2.1 The segmentation of the skier population

We have defined a synthetic population of skiers for the multi-agent model. The skiers are aggregated in groups in order to simplify the simulation, $g \in K$, where $K$ is the set of simulated skiers. These groups are families or friends, for example, and we assume they never separate throughout the simulation.

The population of skiers is divided according to group size $s_{g,g \in K} \in S$, where $s$ is one size among the set of different groups defined, $S$. A homogeneous level of skiing abilities, $l_{g,g \in K} \in L$, is introduced into the group, where $L$ containing all the ability steps.

Moreover, because of the non-separate assumption, group speed $v_g$ is the same for all skiers in a given group $g \in K$ and is defined by the group level and the slope level. The GPS data will provide an accurate assessment of speed. We have also assumed that the number of skiers on the slope does not affect the speed calculation. Details of the speed formulation are provided in a further section, in the part covering calibration.

## 2.3 Behavior modelling

Groups of skiers enter the ski-area at a specific point in the network to which they are attached, so that they can return at the end of their trip. Most of the time, the skiers move along an arc of the network at a speed that is either fixed by technical characteristics if they are on a lift, or that corresponds to their own skiing speed if they are on a slope. Their speed on a slope, $v_{g,l}$, depends on their ability level and the difficulty of the slope. This speed is then assumed to be fixed for each group on each type of track. At the bottom of an ascent, if the demand is higher than the flow proposed on the ascent, a waiting mechanism is created. The group of skiers can also take a break for lunch. This break has a start and a duration (which can be nil, if the group does not take a break) that is specific to each group. It is also assumed that the group does not take any other significant breaks during the day. Small breaks on the tracks are included in the calculation of the average speed, explained in a later section. The core of the intelligence of the model is in the choice that the groups of skiers make at each intersection in the graph. Behavioral patterns are described in Poulhès and Mirial (2017). Below, we will describe the enriched choice model that we have developed.

---

[1] A comfort variable may be introduced in order to distinguish the chair ski-lift without weather protection, cabin cab with or without seat places.

Two types of behavior guide the group. It is either free to go where it wants and does not set itself any constraints, in which case it is called free choice, or it has a very specific objective, which is called directed choice. As soon as it enters the field, the group is given a destination. A parameter $\theta_{l,l \in L}$ defines the time spent in free mode before choosing a new destination.

### 2.3.1 Free choice

When in free choice mode, at each choice point $c \in N$ in the network, the group of skiers must choose a direction among the finite number of slopes $P_c \in P$ and ski lifts $R_c \in R$, arcs in which $c$ is the head node. The probability of each direction is calculated by a logit model (Ben-Akiva and Bierlaire 1999). The utility $U_i$ or the interest is associated with each option $i$ ($i = \{p \in P_c\}$ or $\{r \in R_c\}$) available. The utility consists of a deterministic term $V_i$ and a stochastic term $\varepsilon_i$ : $U(i) = V_i + \varepsilon_i$.

The points below describe the components of the deterministic term for the slopes and the ski lifts.

This first terms (i), (ii), (iii), (iv) and (v) focus only on local choice. The other terms gives direction preferences.

(i) The skiers remember what slope or lift they have already been on and tend to prefer to try new slopes. A negative factor $f_i$ is applied to the lift or slope that the skiers have already been on, which gives an advantage those that have been least explored.
(ii) The difference between the departure altitude $Al_r^-$ of the ski lift or slope and its arrival altitude $Al_r^+$. This means that a ski-lift can be as competitive as a slope. We will represent this term as $a_i$ in $a_i = Al_r^+ - Al_r^-$
(iii) The travel time $t_i$ given for the ski-lifts and calculated a priori according to slope length and group speed on this type of slope.

Specific terms influence the choice of a ski-lift option $i, i = \{r \in R_c\}$:

(iv) The expected waiting time $w_i$ calculated by a specific sub-model. To avoid feedback and interdependence between the variables, we use the waiting time calculated at the previous simulation step time.
(v) The enjoyment $j_{i,i=r}$, which includes comfort and difficulty according to the ability level of the group. For example, there are some difficult ski lifts that beginners avoid. However, speed is taken into account in terms of ski lift type: the groups are not aware of lift speed but they have a ranking of speeds, since a detachable chairlift is much faster than a conventional chairlift.

As with ski-lift utility, slope utility consists of a specific enjoyment term $j_{i,,i=p}$ that includes a number of factors: enjoyment depending on static slope features — (a) length, (b) snow cover (c) view — and dynamic features: (d) sun exposure and (e) skier density.

(vi) For most skiers, the number of slopes in a specific zone of the ski-area influences the group's choice. In other words, in the case of a choice between a ski lift that leads to just one slope and a ski lift that is connected to the rest of the ski-area, most skiers tend to choose the second option. A matrix of minimal distance from each node to all nodes provides an access distance index: $u_i$. $u_i$ is calculated at the tail-node $n_i$ of the arc option $i$ as a ratio. Considering $d_{ii'}$ representing the distance from the node $n_i$ to another node $n_{i'}$ of the network and $D_{ii'} = \sum_{n_{i'} \in N} d_{ii'}$, we define:

$$u_i = \frac{\max(D_{i'}, i' \in N) - D_i}{\max(D_{i'}, i' \in N) - \min(D_{i'}, i' \in N)} \quad (1)$$

The index is close to zero when the distance to other nodes is large and close to 1 otherwise.

This matrix is calculated before the simulation, on the assumption that the skiers cannot have advance information about waiting time. As the upper-level destination choice, this term is part of the skiers' overall vision.

(vii) A last term takes into account the fact that a group will not take a lift that leads only to slopes that are too difficult for its ability level or lacking in interest. Similarly, a series of trails that lead to lifts with little use makes such a route less interesting. The purpose of this term, therefore, is to anticipate both the attractiveness and the feasibility of the route for groups of skiers. For the choice of a slope $p \in P_c$, on the tree of slopes connected to $p$ without passing through a lift, $\bar{r}(p)$ denotes the lift that is directly accessible from the tree with the highest enjoyment level. This value will be assigned to the slope option $p$, $\bar{j}_p = j(\bar{r}(p))$. Symmetrically, we try to compare the attractiveness of a lift with the total enjoyment of the slopes to which it provides direct access. A lift $r \in R_c$ will therefore be given a value $\bar{j}_r = \sum_{p_r} j(p_r)$, where $p_r$ denotes all the slopes accessible from the lift r.

To summarize, the total utility for the option $i = (\{p \in P_c\} \text{ or } \{r \in R_c\})$ for a group of skiers, $g \in K$, , is:

$$U_i = \beta_{(p,r),s_g} a_i - \mathbf{1}_{i=p} \rho_{p,s_g} f_i - \vartheta_{(p,r),s_g} j_i - \theta_{(p,r),s_g} t_i \quad (2)$$
$$- \mathbf{1}_{i=r} \delta_{r,s_g} w_i + \alpha_{(p,r),s_g} u_i + \mu_{(p,r),s_g} \bar{J}_i + \varepsilon_i$$

Where $(\mu_{(p,r),s_g}, \alpha_{(p,r),s_g}, \beta_{(p,r),s_g}, \rho_{p,s_g}, \vartheta_{r,s_g}, \theta_{(p,r),s_g}) \in [0;1]$ is the set of normalized parameters depending on the type of arc (slope or ski-lift) $\{p, r\}$ and the level of the group of skiers, $s_g \in S$.

The multinomial logit[2] model with a Gumbel distribution provides the probability for an option $i$ to be chosen among $C_s$ options with $\omega \in R$, the distribution parameter:

$$P(i) = \frac{e^{\omega U_i}}{\sum_{j \in C_s} e^{\omega U_j}} \quad (3)$$

### 2.3.2 Lift waiting time

The groups of skiers arriving in the lift line are stored on top of the stack of waiting skiers, following a FIFO logic. The first skiers in the stack enter the lift as seats arrive (e.g. chairlift seats or gondola cabin). If the group is larger than the available capacity, the group splits up by filling the first component of the lift as a priority. If there are still places available in the first element to leave, other groups can fill them provided they do not separate. We make the assumption that if the lift is of a kind that allows a whole group to fit into one of its elements, the group will prefer to wait for an entire element rather than to complete an element that is already partially filled. Field observations and comparisons between the maximum numbers of pass swipes and the capacities of the lifts confirm these assumptions that most groups do not split.

---

[2] A nested logit with a first step for the choice between slopes and ski lift is not relevant. In fact, the skier chooses an itinerary. The choice of slope or ski lift is secondary.

### 2.3.3 Destination choice

Every node in the network is potentially a destination. A weight $\varphi_{n, n \in N}$ is associated with the node to calculate the probability of its being chosen. This weight depends on the ability level of the group $l_{g, g \in K} \in L$, since a group of beginners will not aim to reach a node that serves only black slopes.

This weight is calculated as follows from 2 terms, one concerning accessibility in ski lifts and the other the interest for skiers:

$$\varphi_n = \sum_{r \to n} a_r . v_r + \sum_{p \leftarrow n} \gamma_{pl_g} . l_p$$

The first term, lift accessibility, is the sum for all lifts arriving at node n of the product of the difference in altitude with the lift flow. The second term is simply the total length of runs $l_p$ accessible from the node weighted for each run by a parameter $\gamma_{pl_g}$, which gives the interest of each type of group in each level of run.

### 2.3.4 Guided choice

The possible types of objective are as follows:

(i) To join a specific node in the ski-area that the skier has chosen to suit his or her own characteristics.
(ii) To take a lunch break, i.e. choose a place to stop and reach there.
(iii) To return to the point of entry into the ski-area

In cases (i) and (ii), the $g \in K$ group seeks to reach its destination by following the shortest route in the shortest amount of time at the time it makes its choice. In the case of (iii), return to the station, the choice is not totally directed but partially constrained. From a certain time $H_g^d$, for each possible choice, the group calculates the shortest time to return to its point of origin before the ski-area closing time $H^S$ with a safety margin $h^S$. If this choice does not make it possible to return before the area closes ($H^S - h^S$), the group does not choose this option. The safety margin takes into account the uncertainty of lift waiting times, which can be significant on some return routes to the resort.

A piece of software named Dynaski was coded in java with a graphical user interface. The JUNG library, an open-source library,[3] was used to enable users to code the network interface.

As the case study illustrating the model is the La Plagne ski-area, all the elements presented in chapter 3 correspond to the calibration values specific to the La Plagne area. However, the methodology is general and can be adapted to any ski-area.

## 3 Calibration method

The model must correspond to the specific behaviors of the skiers in the ski-area being studied. For this, the model can be calibrated at 3 levels: (i) A skier population corresponding to a typical day of the simulation, (ii) For each group type defined by ability level, an interpolation of the average speeds by slope type is performed from the GPS data collected with the ski-area app, (iii) The parameters of the slope and lift utility functions per group type estimated by minimizing the error between the simulation results and the lift pass data.

---
[3] http://jung.sourceforge.net/

In order to adapt to the specificity of a ski-area, the data needed and available for the area are specified for each calibration level: (i) The composition of the skier population must be reset for each ski-area modelled, although a generic population can be used. (ii) Few ski-areas have their own apps. Furthermore, the speed per skiing ability level and slope level can be considered identical regardless of the area simulated, and it is only the distribution of the number of skiers per level that will change. (iii) All the major ski-areas in the world have "turnstiles" and collect turnstile pass data that model can use to match the specific behaviors of the skiers in the area.

### 3.1 Group characteristics from survey

A survey was carried out in the La Plagne ski-area in February 2018, with 630 groups responding to the interviewers' questions. The survey had to be carried out on the same area and with the same type of population, in this case, one week of school holidays in February. This type of survey has never been done in La Plagne or remains confidential for marketing reasons, which is why we have no benchmark to compare and check whether our sample of respondents is representative.

In this case, the survey has 2 main purposes. The first is to limit the value intervals for certain parameters in the model, in particular the intervals for the start of the lunch break and lunch break duration, as well as the other break times during the day. The second is to build a synthetic population. That is what will be clarified here.

The total population using the ski-area is defined per station from the count data. But this population is not characterized and simply records the total number of people. The model, however, takes into account groups of skiers who each have their own ability level, which also determines their behavior in terms of their preference for certain slopes and speed. The characteristics of the groups also depend on the station of residence, a finding supported by a question in the survey. However, it is not possible to construct a population based on levels, group size and station of origin from the sample of respondents. We therefore chose to construct a population that does not depend on the resort of origin and is therefore homogeneous over the whole ski-area.

In any group of skiers, there are obviously disparities in ability levels and each group, even with identical characteristics, has preferences for a certain specific type of slope. The survey asks several questions that can be used to build a population based on practices:

- How many people are currently in your group?
- Does everyone in the group ski/snowboard at the same level?
- What is the group's average level? / What are the group's minimum and maximum levels? (2 answers needed)
- What types of slopes do you prefer to go on? (Multiple answers allowed)
- If you ski off-piste, what percentage of your time do you spend doing this (between 0 and 100)?

From the raw questionnaire results, we obtain the group size distribution shown in the histogram in Figure 1. The average group size was 5.5 skiers during the vacation week in which the survey was conducted. This may be explained by the presence of large groups such as summer camps or student trips.

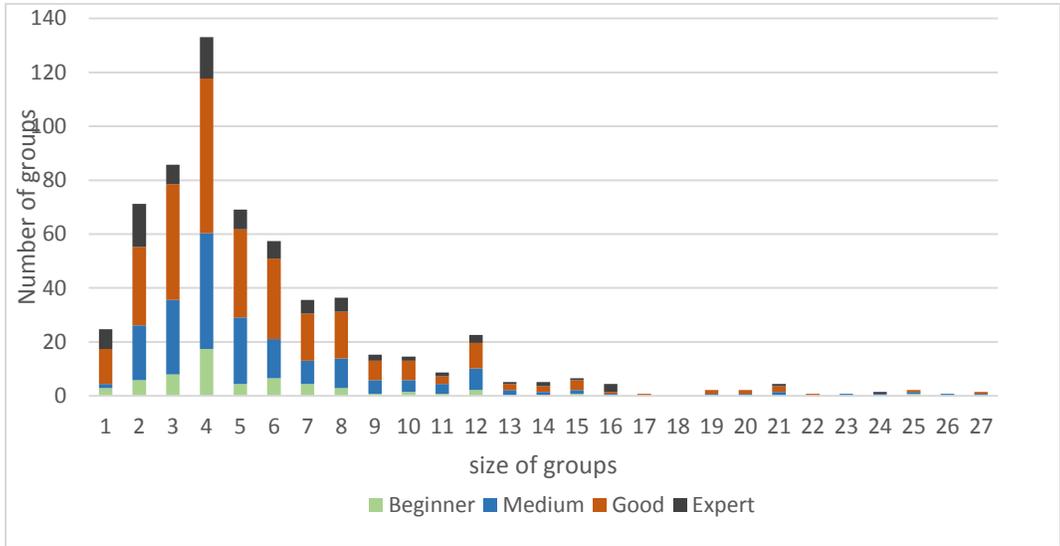

Figure 1: Distribution of group sizes of La Plagne skiers according to the February 2017 survey

In order to ensure that the synthetic population is statistically representative, certain group sizes are aggregated and levels are standardized. It is noted that groups with disparate levels also have specific practices, especially groups with beginners. Considering that there are few groups of pure beginners, we simplify the number of groups by distributing all beginners in the same group but give them practices that are the average of these groups.

In the end, by weighting the groups by the number of skiers, we define the distribution presented in the table 1.

| Size of group | Beginner | Medium | Good | Expert |
|---|---|---|---|---|
| 2 | 1.3% | 3.7% | 6.3% | 3.5% |
| 4 | 4.1% | 11.3% | 15.8% | 3.7% |
| 6 | 1.8% | 6.2% | 10.1% | 2.2% |
| 8 | 1.2% | 3.3% | 5.8% | 1.7% |
| 12 | 0.8% | 3.3% | 4.2% | 1.3% |
| 16 | 0.4% | 3.0% | 3.6% | 1.4% |

Table 1: Distribution of the synthetic population of the ski-area according to group size and level

### 3.2 Speed calculation from GPS-Data

The company that operates the Les Arcs ski-area employs the YUGE app, which is available to skiers in the Paradiski area (Les Arcs and La Plagne). We use it to estimate skiers' speed distributions according to the type of slope. The objective is to build a speed distribution per skier level and per slope type.

The data provided are not associated with skier profiles or slopes. The first step, therefore, is to associate the GPS tracks with the geographical layer of the slopes in the ski-area, allowing a margin of error of 30 m. We then apply the frequency of use of the track types by the skiers in the survey presented in the previous chapter. Assuming that the survey populations and the GPS base are identical (they correspond to the same week), we distribute the tracks according to the time spent on each type of slope. This gives average speed distributions per slope and per skier level. These speeds take into account break times on the slope, but not at the bottom of slopes. This is because the areas at the base of the slopes are common to the different ski lifts and are therefore not identifiable: we cannot distinguish between a person waiting in line for a lift from a person waiting for a member of his group at the bottom of the slope. The normal law best represents these distributions.

We were provided with the GPS data for February 22, 2018 in the La Plagne ski-area. The weather conditions on that day were good, with good visibility. The speed profiles were those of a normal ski day, when the traffic is greatest. The data contain 1986 identifiers corresponding to 934,089 valid records, i.e. with no inconsistencies in the values (i.e. 84% of the raw data). A record is a GPS coordinate associated with a capture schedule. In order to avoid over-estimated values, we removed from our sample the skiers who spent less than 30 minutes on the slopes. This left 1259 skiers for analysis. Figure 2 illustrates the distribution of the skiers' GPS points in relation to the color of the slopes they skied. The green and black slopes are the least represented and the estimated distributions will therefore be the least reliable.

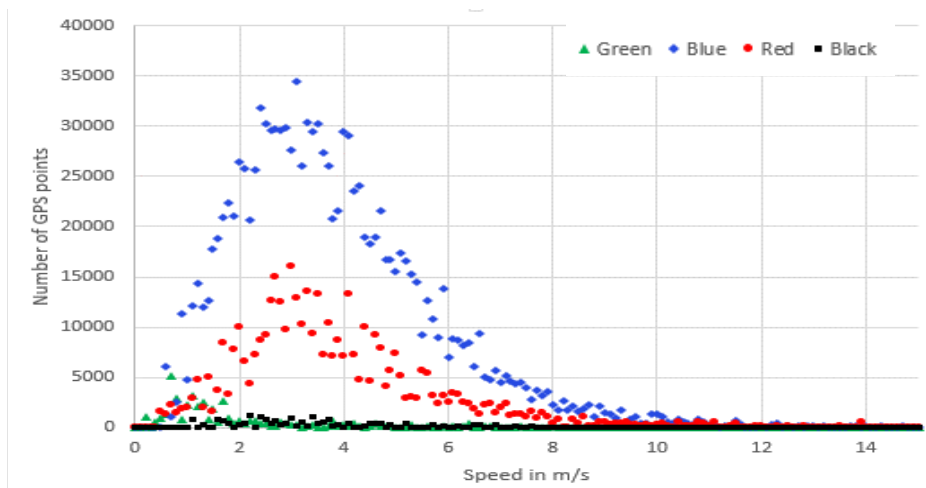

Figure 2: GPS point distribution according to speed per slope level.

The normal distribution laws $\mathcal{N}(\mu_l, \sigma_l)$, $l \in L$, have the values shown in Table 3 and the profiles shown in Figure 3.

During the GPS recording day, the time spent on the red and black tracks by an equivalent proportion of beginners is negligible. For this reason we could not calculate a speed distribution for the red and black beginners' tracks. The same is true for the means on the black tracks. However, the survey data specified that whatever the ability level of the group of skiers, all types of runs were covered. We can assume that the recordings for a single day do not allow us to account for the diversity of each individual's practices, but we can also assume that the groups may give false information and that the beginners claim to cover ski on red and black runs when they do not.

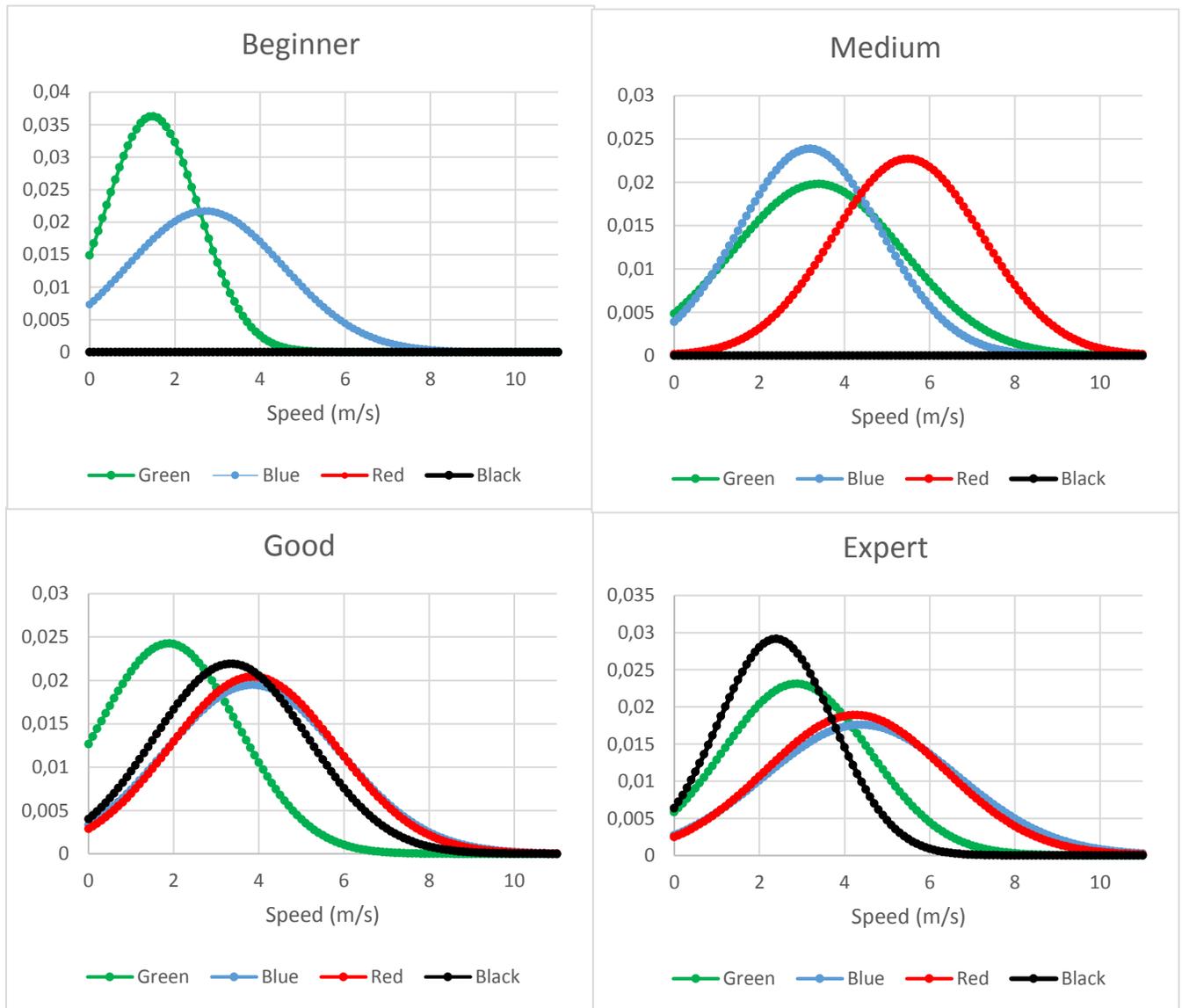

Figure 3: Speed distribution according to slope levels

|  |  | Green | Blue | Red | Black |
|---|---|---|---|---|---|
| **Beginner** | Mean | 1.47 | 2.72 | | |
| | Standard deviation | 1.21 | 3.39 | | |
| **Medium** | Mean | 3.38 | 3.18 | 5.49 | |
| | Standard deviation | 4.06 | 2.79 | 3.08 | |
| **Good** | Mean | 1.88 | 3.84 | 3.86 | 3.35 |
| | Standard deviation | 2.70 | 4.19 | 3.81 | 3.31 |
| **Expert** | Mean | 2.87 | 4.39 | 4.26 | 2.39 |
| | Standard deviation | 2.99 | 5.15 | 4.46 | 1.87 |

Table 3: For each slope level-ski level pair, the parameters of the normal law of speed distribution.

## 3.3 Calibration from the ski pass data

The Automatic Fare Collection (AFC) provided by the La Plagne ski-area corresponds to the French winter school vacation in 2014. As France is divided into 3 zones which have 15 days of school vacation, each a week apart, this period corresponds to 1 month of vacation, so the 3 zones are never on vacation at the same time.

Let t be the time interval that corresponds to the time step in which the pass data per ski lift are provided, and $T$ be all the time steps in the simulation day. $N_s(r, t)$ is the number of skiers who swiped their passes during period t at lift $r$.

The AFC for La Plagne are available with a time step of 30min. The sum of the swipes per day during the whole holiday period is shown in figure 4. The days when the number of swipes is very low (15,000 – 20,000) correspond to Saturdays, the day on which the rental period begins and ends for the vast majority of accommodation in the resort. These days are therefore very specific, as the skiers tend to be locals who come for the weekend. Small decreases in attendance such as day 4 or day 20 in 2015 correspond to bad weather days when weather conditions made the ski-area less attractive and skiers chose other activities.

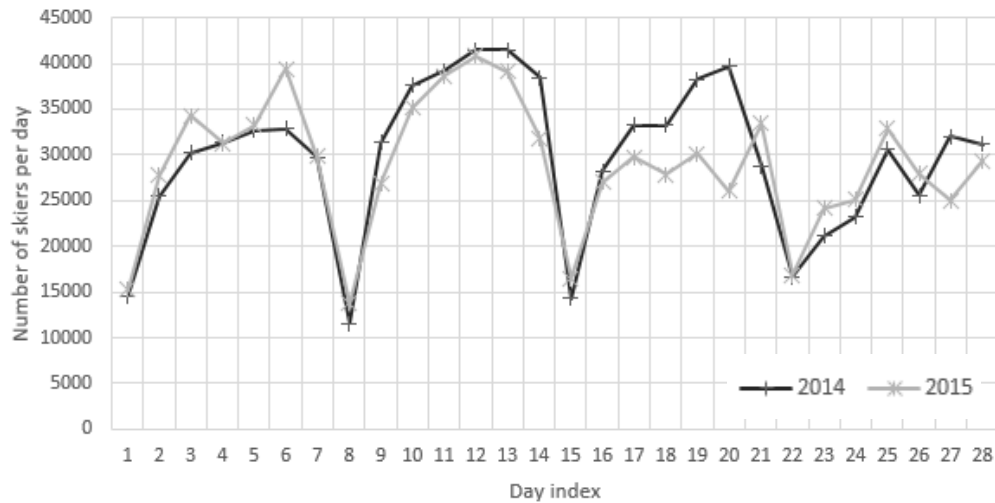

Figure 4: Profile of the number of validations per day over the school holiday months 2014 and 2015

Figure 5 shows for a very busy lift in the area, the Blanchet detachable chairlift, the daily temporal evolution of the number of validations per ½ hour on average for the 24 days (excluding Saturdays) of the French school holiday period of 2014. The standard deviation is also shown for each mean value. A certain homogeneity in the number of validations can be observed, especially during the busiest periods when the maximum capacity of the upwelling unifies the flows. These periods correspond to mid-morning and mid-afternoon. The validation data also confirm the discrepancy between the theoretical flow rate, which is the number of available seats per chairlift times the number of chairlifts in a time period, and the actual flow rate. This confirms the difficulty of the groups to separate and for this 6-seater chairlift, for each row, one seat and empty on average.

This lift is also very quickly very much used in the morning until it closes. This corresponds to the statements of the respondents who inform for the most part (80%) arriving between 9am and 10am and 82% leaving the area between 4pm and 5pm.

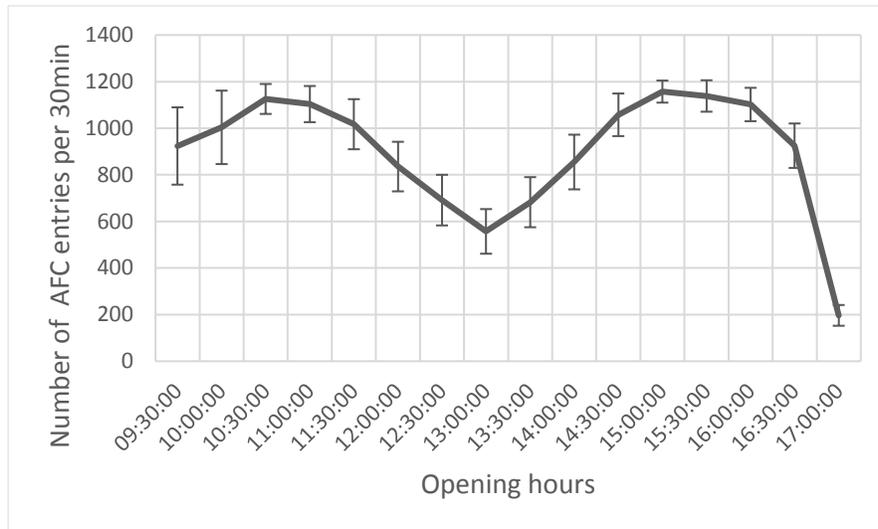

Figure 5: Average number of swipes per 30 minutes for the 28 days of the base in 2014 for a detachable chairlift with a theoretical capacity of 1500 skiers/30min.

The parameter of the discrete choice model and the parameters of the utilities per slope and lift are calculated in a classical optimization problem. The objective is that the simulation results should be as close as possible to the swipe results. A heuristic optimization method is used to get closer to the solution that minimizes the error.

# 4 Demonstration on a large-scale ski-area

## 4.1 La Plagne ski-area

La Plagne is one of the largest integrated resorts in France. With 2.5M skier-days in 2018, it is the world's leading resort in terms of skier numbers. This corresponds to an offer of 65,000 beds in 9 "villages". The associated ski-area consists of 225km of slopes and 94 RM (data from the Ski lift company: Société d'Aménagement de La Plagne, SAP). Since the construction of the Vanoise Express cable car in 2003, the area has been connected to Les Arcs ski-area, offering skiers a total of 450km of pistes, across a connected ski-area called Paradiski.

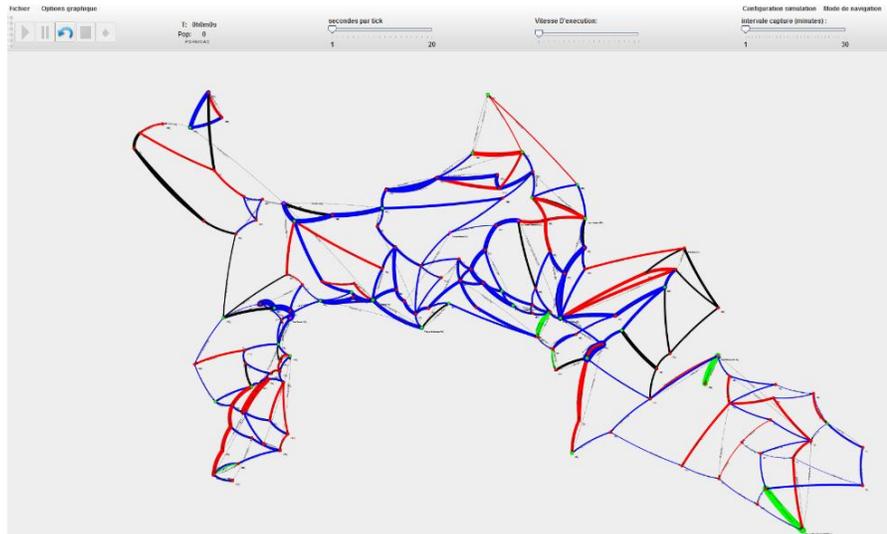

Figure 6: Screenshot of the Dynaski graphical interface illustrating the ski-area of La Plagne

## 4.2 Data

A certain quantity of data specific to the La Plagne ski-area were needed to perform the simulations and were provided by the operating company. The slopes and lift network graph can be constructed from maps of the ski-area and a digital elevation model to have the intermediate altitudes of the slopes that are to be divided.

The ski resorts and parking lots connected with the ski-area are the nodes that generate skiers. For each generator, a node representing the arrival of skiers in the area is created at the center of gravity of the location of the skiers it represents. In order for the skiers to get to the slopes, connectors that link this virtual node to the rest of the network are also added to the ski-area graph.

In order to estimate initial skier demand, we use the first swipe data which gives the number of swipes recorded per lift and per time slot for each skier just for the first use of the day. This demand is aggregated for each ski resort represented by the generating nodes, which are unique as described above. Swipes per lift are allocated to the nearest ski resort. The operator can also define car parks as demand generators or disaggregate the resort into several districts.

For the La Plagne ski-area, the flows of skiers in 30-minute steps at the lifts are aggregated to each resort or car park connected to the La Plagne ski-area. A Poisson distribution law is used to simulate a random distribution of skiers arrivals in the area in time steps.

## 4.3 Validation data vs simulation results

The survey of skiers in La Plagne during a school vacation week was used to construct a population representing La Plagne during the simulation period corresponding to the swipe data provided by the Ski lift company. The distribution of skiers' speeds was constructed from the GPS tracks of skiers in the La Plagne ski-area.

The simulation day is February 28, 2014, a day with a large number of skiers. As explained in Doctor and Scaglione (2007), days with congestion are days of good weather. Optimization is based on swipe data provided by the Ski lift company. To limit the impact of randomness in the simulation results, 10 runs of each scenario were performed. The results shown are the average of these 10 runs.

In the simulation, the number of skiers injected into the ski-area corresponds to the number of first swipes, i.e. 40,167. By counting the number of total swipes recorded during the day at the lifts and by counting the number of skiers who went to the lifts in the simulation, we can compare the average number of lifts per skier. The resulting values are shown in Table 4.

| **Simulation count** | 7.36 Lifts/skier |
|---|---|
| **Swipe count** | 7.37 Lifts/skier |

Table 4: Comparison between swipe data and simulation: number of lifts per skier on the simulation day

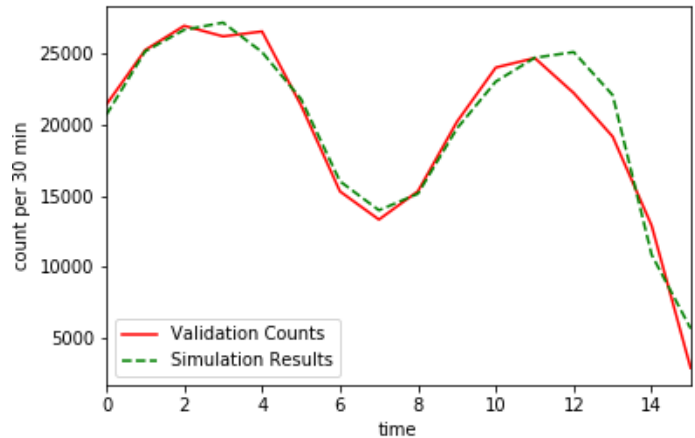

Figure 7: Daily comparison between the total average number of swipes per 30 minutes over 1 month of school vacation in 2014 and simulation results.

Figure 7 shows the comparison between the 30-minute swipe counts (in red) and the simulation results during the day (in green). The morning load and the drop in ski-area attendance at noon are fairly well simulated. The overall results at the end of the day are less accurate and reflect the difficulty of returning to the resort. In the simulation, the skiers are more spread across the ski-area than in reality and therefore take longer to return to their point of origin.

### 4.4 Case study: extension of a resort, creation of accommodation for skiers

A prospective use of the tool illustrates the Dynaski model through a real case, the construction of new tourist residences in one of the La Plagne resorts. The comparison of a scenario with and without this development project provides some guidance to the impact of new residences on congestion in the ski-area.

A major housing construction project is planned at the La Plagne 2000 station, which would double the current number of beds (Antoine 2017) by adding 2450 beds. In order to assess the impact of the construction of new stations or new housing, the new demand generated needs to be estimated. A passage coefficient will therefore need to be constructed between the number of beds planned and the number of skiers generated.

The La Plagne high altitude resorts have 43,725 residential beds. Dividing the number of first swipes on a school vacation day by the number of beds gives an estimated ratio of 0.66 skiers/bed. This results in 1620 new skiers being generated for injection into the Aime 2000 node and by approximation, following the same 30-minute entry distribution as the other Aime 2000 skiers. This represents a 3.5% increase in attendance over the entire ski-area.

The simulation results for the ski day of February 28, 2014 are compared with the projections for the same day after the urbanization project, all other things being equal.

The first important result is a modest increase in waiting times. The average wait times at the lifts are 5.4 min without the project and 5.6 min with the urban project. Table 6 shows the distribution of skiers' times during their presence in the ski-area. The addition of 1620 skiers to the ski-area does not change the time spent on the slopes by all skiers. On the other hand, waiting time increases by 12% and time in the ski-area decreases proportionally. The 3.5% increase in the number of skiers in the ski-area does not significantly alter the time spent per activity. The differences calculated are within the model's uncertainty margins and therefore cannot be considered robust. However, it can be concluded that the increase in waiting time at the ski lifts would not be significant enough to create a risk of congestion that would be detrimental to the attractiveness of the ski-area.

|  | **Skiing time** | **Waiting time** | **In ski lift time** |
| --- | --- | --- | --- |
| **Base scenario** | 65% | 16.5% | 18,5% |
| **With urban project** | 65% | 18.5% | 16,5% |

Table 6: Distribution of ratio of skier time in ski time/waiting time/ ski lift time

Other results complete the analysis. Figure 8 shows time changes with the urban project in the major ski-area lifts or those with significant waiting time in the simulation. It may be observed that the increase in waiting time is not homogeneous across the ski-area and that some lifts show a decrease in waiting time (Belle Plagne, Montchavin) while others show a sharp increase in waiting time (Roche or Colosses).

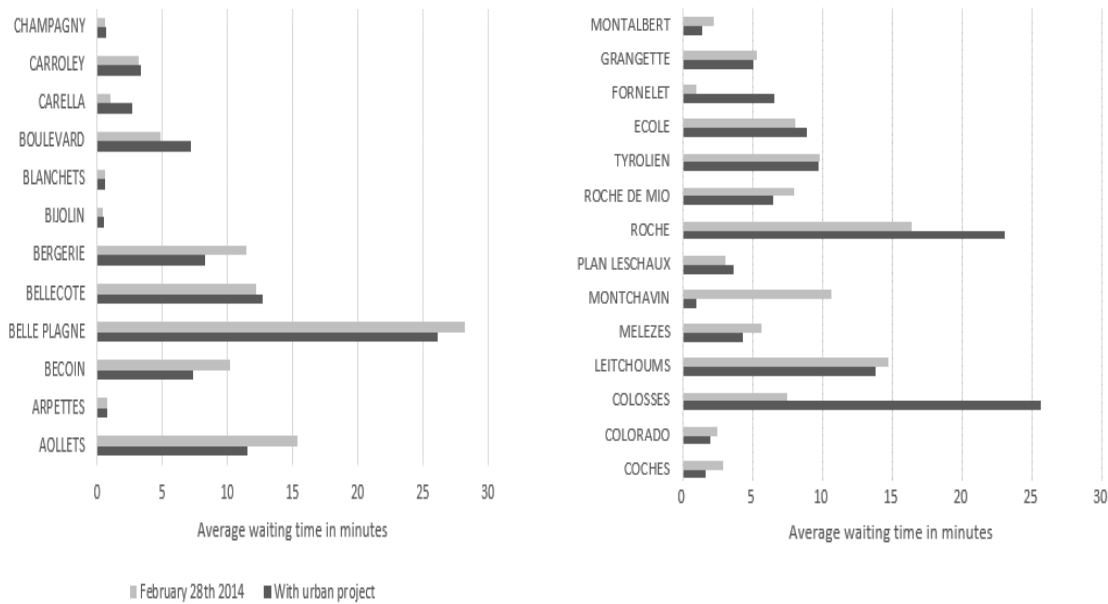

Figure 8: Comparison of average waiting time per main ski-lift

Original indicators used today by ski-area companies, which have no reference, can be obtained to test various projects with the Dynaski simulator. The objective of this very simplified study is not to reach an accurate assessment of an urbanization scenario but to demonstrate the possible use and contribution of this type of tool in the evaluation of ski station projects, the first of its kind.

## 4.5 Discussion

There are several major uncertainties and limitations in the model. The first concerns the behavioral model. Because of a lack of data and knowledge, only one type of behavior has been modelled. These behavioral choices correspond either to elements tested in the surveys as the importance of waiting time or to structuring but intrinsic elements in the choices as the interest in gaining altitude. The survey results also show that most skiers move around the ski-area without thinking about their destination in advance. However, simulations with 100% random behavior show that most skiers find themselves at a certain point in the simulation in a sub-graph of the network that they cannot easily leave without knowledge of the ski-area and therefore a strategy in behavioral choices. Skiers therefore have to choose destinations at certain times or in certain zones of the ski-area. The choice of these destinations could be better understood from targeted surveys and the analysis of skiers' trajectories around the ski-area.

The original databases used for model calibration are classic data for urban transport models. Population census and household-travel surveys are used to characterize traveling populations and to construct a synthetic population (Rich and Mulalic, 2012; Zhang et al., 2019). The survey of the population of skiers in the domain is the first one whose primary objective is to qualify this population with regard to these mobility practices in the ski-area. The populations of skiers change over the course of the season within the same ski-area, according to whether a period falls within or outside school vacation times, or in another holiday season. But for our study which is limited to one week of specific school holidays in the year and the consistency between the data over time limits this type of error. Perhaps there are certain population profiles that are conducive to congestion and long waiting times at the lifts. An in-depth study of skier populations and their behaviors would make it possible to apply the simulation tool over the whole season, and especially to compare the impact of population types on levels of fluidity in the ski-area.

GPS data or swipe data recorded for each individual would be natural data sources for this. This research on destination choices in the ski-area is comparable with research on attraction choices in theme parks and the simulation of visitors between attractions. As identified by Kemperman et al. (2004) in theme park, the first-time skiers and the repeaters have different behaviors. However, most of skiers stay 6 days in the same ski-area during their holidays, hence they become repeaters and the number of first-time skiers differ from the Sunday to the Friday in a same week. If the individual behavior changes, the number of validations per lift is constant in the week.

GPS data are used in many research for tracking tourist movement (Shoval and Isaacson, 2007) and particularly in the theme park, studies use GPS data to better understand the visitor behaviors (Biremboim et al., 2013). In this research, GPS data are used to estimate the skier speed. But, they can improve the break time estimation. However, if skiers take a break at the top of the lift or in the area at the bottom of the lift or in an area that cannot be considered as a slope, this break time will be difficult to estimate automatically. The speed of the skiers, or even the time they spend on the slopes, vary according to their origins, their knowledge of the ski-area, and their choice behaviors. This new dynamic GPS data on smartphones is beginning to provide spatio-temporal speed maps for both road vehicles (Yu et al., 2020) with fine vehicle detection that finely informs congestion areas. Or for bicycles (Clarry et al., 2019) where speeds make it possible to build a statistical speed model as a function of the urban environment. Additional work on a much larger volume of data or more precise field surveys on stops in the ski-area would provide a better understanding of the break time, which is largely uncertain in this study. More consistent GPS data would also permit an in-depth analysis of how speed on the slopes is affected by skier density, adding precision to the speed estimates, which would no longer depend only on ability levels and slope type, but would also include congestion on the slopes. Average gradient is also a determining factor for speed, which is only taken into account in the slope color.

AFC data are also fairly recent and are mostly available and used in research to better understand mobility behaviours in transit network (Gordon et al., 2018). Rahbar et al. (2020) use one-day AFC data to calibrate the assignment results of a model according to the time period. The following day they use them to validate the results, obtaining errors of less than 30% on the rail transit and more than 30% on the buses. Differences in flows and ski conditions between consecutive days make this calibration and validation approach difficult in the case of ski areas. The results of the calibration we obtain are encouraging but will need to be validated over another year to check whether the estimated parameters remain valid for other initial conditions.

Several questions arise about the input data that have so far not been answered. These input data allow us to refine the simulation model by proposing results that are in principle better than they would be otherwise. However, each of these pieces of estimated data introduces a margin of error. A sensitivity analysis of the input data thus constructed will have to be undertaken on the La Plagne ski-area and on other ski-areas. Answers can then be given to the following questions: can these data be used as they are on another ski-area or do they have to be reconstructed? If the sensitivity of the results to this input data is low, perhaps it is not essential for a particular use of the model. Conversely, if a sensitivity analysis reveals significant gap, then the input data should be further investigated.

# Conclusion

After explaining the operational interest of such a model, we presented the construction of a multi-agent model for the simulation of skiers in a ski-area. Since this field of research has received little scholarly attention, certain data needed for the model had to be constructed. A field survey provided a better understanding of the mobility behaviors of the skiers and also the distribution of skiers in groups of ability levels and size. GPS data were used to estimate the speed distributions of skiers on the slopes, which are not available in the literature today. Finally, the lift swipe data are used as an input to estimate skier demand on a typical simulation day, but also as calibration data for the model parameters. An application to the large ski-area of La Plagne illustrates the model and demonstrates its operationality for most ski-areas around the world. Such a model offers a wide variety of uses and is relevant for the actors with a professional involvement in winter sports. Nevertheless, the results obtained need to be confirmed with more consistent data and operational projects that have already been implemented. This article open a new companion research field of the theme park with a lot of specificities. This planning tool can also be adapted for other leisure places and theme park.